**Selective Reflection Spectroscopy on the UV Third Resonance Line of Cs :**

**Simultaneous Probing of a van der Waals Atom-Surface Interaction Sensitive to Far IR**

**Couplings and of Interatomic Collisions**

Pedro Chaves de Souza Segundo, Ismahène Hamdi,

Michèle Fichet, Daniel Bloch and Martial Ducloy

*Laboratoire de Physique des Lasers, UMR7538 du CNRS et de l'Université Paris13*
*99 Av JB Clément, 93430 Villetaneuse, France*
*e-mail : bloch@lpl.univ-paris13.fr*

**Abstract** : *We report on the analysis of FM selective reflection experiments on the $6S_{1/2} \rightarrow 8P_{3/2}$ transition of Cs at 388 nm, and on the measurement of the surface van der Waals interaction exerted by a sapphire interface on $Cs(8P_{3/2})$. Various improvements in the systematic fitting of the experiments have permitted to supersede the major difficulty of a severe overlap of the hyperfine components, originating on the one hand in a relatively small natural structure, and on the other hand on a large pressure broadening imposed by the high atomic density needed for the observation of selective reflection on a weak transition. The strength of the van der Waals surface interaction is evaluated to be $73 \pm 10$ kHz.$\mu m^3$. An evaluation of the pressure shift of the transition is also provided as a by-product of the measurement. We finally discuss the significance of an apparent disagreement between the experimental measurement of the surface interaction, and the theoretical value calculated for an electromagnetic vacuum at a null temperature. The possible influence of the thermal excitation of the surface is evoked, because, the dominant contributions to the vW interaction for $Cs(8P_{3/2})$ lie in the far infrared range.*

**1. Introduction**

The long-range atom-surface attraction is a fundamental effect that ultimately explains the cohesion of the matter. If the experimental evaluations of this interaction remain scarce with respect to its universality (see *e.g.*[1] and references therein) and to the extremely broad range of energy it spans over, available measurements have confirmed a reasonable agreement with the theoretical predictions. In the principle, the accuracy of the theory is potentially very high, as simply dependent upon the determinations of atomic wavefunctions, that have been precisely known for a long time as far as alkali-metal are concerned. If recent fascinating advances were recently demonstrated in the related Casimir interaction between two surfaces [2], the near-field regime for which the retardation effects can be neglected permits to enter into the frame of the electrostatic van der Waals (vW) surface interaction [3-5], allowing to simply consider an interaction between the fluctuating atom dipole and its induced electrostatic image. This approach implicitly assumes the vacuum field to be at null temperature. However, the atom-surface coupling is known to be strongly sensitive to the virtual dipole couplings to neighbouring energy levels between an atomic level and its neighbours, so that high-lying states are mostly sensitive to a vW interaction whose dominant origin lies in far infrared (IR) couplings [1,6,7]. In such a case, the temperature of the surface, governing at equilibrium the temperature of the vacuum field, may affect the atom-surface interaction [8-11], hence providing a simple example of atomic quantum properties affected by a thermal bath [12]. Such a modification of the atom-surface interaction can be particularly expected when the atomic coupling in the thermal IR range exhibits a resonant coincidence with a thermally populated surface mode.

The present work is devoted to an attempt of a precise experimental evaluation of the surface interaction exerted on a Cs atom in the $8P_{3/2}$ level, a level undergoing a notable (absorption) coupling to Cs(7D) through transitions lying in the thermal infrared: the wavelength of the $8P_{3/2} \rightarrow 7D_{3/2}$ line is 39μm, corresponding to T = 500 K , while the one of the $8P_{3/2} \rightarrow 7D_{5/2}$ line is 36μm corresponding to T = 550 K. The results presented here are obtained at the interface with a sapphire window, a material without surface resonances in this thermal IR range. Although these measurements were acquired for various Cs densities (and vapour temperatures), the variations of the temperature of the window remain negligible. They were mostly intended to provide a reference experiment and a feasibility test in view of planned investigations, on the same atomic level Cs ($8P_{3/2}$), of a severe temperature dependence of the surface interaction [9], when surfaces such as $CaF_2$ or $BaF_2$ exhibit electromagnetic mode resonances in the far thermal IR susceptible to couple with the virtual atomic absorption [10].

Our present experimental evaluation relies on an analysis of the Selective Reflection (SR) spectra obtained at a normal incidence, at a sapphire interface on the third resonance line of Cs ($6S_{1/2} \rightarrow 8P_{3/2}$) at 388 nm. We have developed for a long time linear SR spectroscopy, in order to be able to extract, through an elaborate lineshape analysis [1,8,13-15], the strength $C_3$ of the vW interaction, that shifts the transition energy as $C_3 z^{-3}$ (z: the atom-surface distance). Indeed, SR spectroscopy, which turns to be a Doppler-free technique in its Frequency-Modulated (FM) version [16], typically probes a region $\lambda/2\pi$ away from the surface. The encountered experimental challenges are here comparable to those solved long ago for our first observation of the "strong" van der Waals regime on the (relatively weak) line of second resonance of Cs ($6S_{1/2}$-$7P_{3/2}$, $\lambda$=455 nm)[6,7]. The major difficulties lie in the weakness of the optical transition -the oscillator strength is $\sim 10^{-3}$ -, and in the very narrow hyperfine structure (see fig.1). Hence, high atomic densities must be employed for sensitivity reasons, implying a pressure broadening that tends to make the hyperfine components unresolved. Also, the experiment requires a tuneable UV laser, here provided by a specific diode laser.

## 2. Experimental observations

As usual for such a vW measurement through SR spectroscopy, the set-up (fig. 1) comprises a tuneable FM laser, the SR experiment itself, that requires a cell of heated Cs vapour, and an auxiliary Saturated Absorption (SA) experiment, providing the frequency reference needed to evaluate the vW interaction affecting the SR spectrum.

The laser is a commercial extended cavity diode laser (Toptica DL 100). It includes a grating for line selection in Littrow configuration, and it is equipped with a wavelength-selected semi-conductor chip able to cover the 387.7 nm and 388.9 nm respective lines $6S_{1/2} \rightarrow 8P_{3/2}$ and $6S_{1/2} \rightarrow 8P_{1/2}$. Until now, the experiments have been limited to the $6S_{1/2} \rightarrow 8P_{3/2}$ transition, because the known Cs fine structure anomaly [17] is responsible for a weak oscillator strength of the $6S_{1/2} \rightarrow 8P_{1/2}$ transition - weaker than the $6S_{1/2} \rightarrow 8P_{3/2}$ transition by about an order of magnitude-. The output power does not exceed 3 mW, and drops to ~1mW after optical isolation and some spatial cleaning of the beam structure. The long-term laser jitter, essentially governed by the relative length stability of the extended cavity, is ~4 MHz, twice larger than currently obtained in the near IR with an analogous cavity. The FM (at 12 kHz) is here applied directly through the laser drive current. The laser has the ability to scan more than 10 GHz, notably covering the 9.2 GHz of the Cs hyperfine structure of the ground state. This is obtained however at the expense of an important change (~50 %) in the laser intensity, connected to the near-threshold operation ($I_{op} \leq 44$mA, $I_{th} \approx 39$mA) with an external grating coupling of our specific UV diode laser. Our study actually concentrates on narrow ( $\leq 1$ GHz) regions of the spectra but for long scans, this would imply to renormalize our recordings with the actual output power. Tests of the linearity of the frequency scan with the drive voltage, conducted with the help of a Fabry-Perot analyzer, reveals variations on extended scans (~ 10 GHz) by up to $\pm 10\%$ . This non linearity, attributed to a nonlinearity in the PZT that drives the grating, remains mostly quadratic, making shorter scans more linear, but with a novel calibration (drive voltage *vs*. frequency) required each time that the frequency interval is modified. The vapour cells are T-shaped sealed cells, described elsewhere, with sapphire windows (*c*-axis perpendicularly oriented), that are compatible with a high temperature vapour. The windows are contacted to a metallic tube [6,7] or to a sapphire tube [18]. They are independently heated up in a two-chambers system - main body, and Cs reservoir- , with the Cs reservoir temperature normally governing the atomic density, and the windows slightly overheated (by ~ 20° C) to avoid condensation.

The linear nature of SR spectroscopy makes the (FM) SR signal relatively easily to detect, independently of the incident beam power ($\leq 1$mW), and diameter (~1 mm). Moreover, and in spite of the weakness of the oscillator strength, the low amplitude noise of our UV diode laser makes the SR detection less challenging than with a blue dye laser, such as used in [6,7]. Note that the ultimately attained sensitivity is not exactly as good as for near IR resonance lines, where the photonic shot-noise limit can be approached : this is partly due to the non ideal quantum sensitivity of the photodetector - even for a "UV enhanced" photodiode-, partly to the technical noise of the UV laser, that operates relatively near-threshold. The relative change $\Delta R/R$, in excess of $10^{-4}$ for Cs temperatures typically exceeding 160 °C (see fig.2a) , is conveniently observed through a synchronous detection of the FM provided that the FM amplitude is comparable to the SR signal width. However, as illustrated in fig 2, the required high temperatures and densities ( $\geq 5.10^{14}$ at/cm$^3$) makes the hyperfine structure hardly recognizable, when not totally hindered. Moreover, the combination of dispersive-like shapes (typical of (FM) SR spectroscopy) and of strong vW-induced distortion and broadening, implies a strong overlap between the individual hyperfine components.

In some sense, the auxiliary SA experiment poses more experimental difficulties than the SR set-up itself. In spite of a much longer effective length of interaction (the SA cell is 1cm-long, while the SR experiment typically probes a region ≤100 nm from the interface), the nonlinear nature of the SA experiment, combined with the weak oscillator strength and the limited available power, imposes a strong beam focusing ( ~50 μm diameter, practically limited by the onset of divergence after a Rayleigh length), and subsequent careful pump-probe alignment. To eliminate the residual Doppler-broadened linear absorption appearing in a (FM) SA experiment, a chopper for amplitude modulation (AM) is implemented on the pump beam. A few % of linear absorption, along with narrow SA lineshapes, is obtained for a Cs temperature in the 80-100°C range. In these conditions, the SA lineshapes are hence insensitive to moderate temperature variations, establishing that they provide an adequate frequency reference for the isolated atom transition. Anyhow, one notes obvious imperfections in the SA spectra, that should normally be composed of successive Lorentzians : the relatively long scan time imposed to get acceptable signal-to-noise ratio through the integration time (time constant of the detection : 0.1 s) makes indeed the experiment sensitive to short-term drifts affecting the laser frequency.

The essence of the experiments consists, for various temperature conditions of the SR cell, of the simultaneous recording of the (FM) SR spectrum and of the (AM) SA reference spectrum. As usual in the analysis of SR spectroscopy for vW extraction [6-8], the key-point is to extract a fixed atom-surface interaction in spite of phenomenological changes appearing in the SR lineshapes when modifying the pressure effects (*i.e.* the relative influence of the atom-atom interaction). The spectra were recorded in scans of a limited amplitude (≤ 1 GHz), and repeated to study successively the two hyperfine components $6S_{1/2}(F=3) \rightarrow 8P_{3/2}$ and $6S_{1/2}(F=4) \rightarrow 8P_{3/2}$. In spite of differences in the hyperfine structure of $Cs(8P_{3/2})$, that remains unresolved in the context of SR experiments, the two components F=3→F'=2,3,4 and F=4→F'=3,4,5 normally provide a similar information relatively to the $8P_{3/2}$ level. Indeed, the surface vW interaction is expected to be insensitive to the considered hyperfine component as long as the Zeeman degeneracy is not removed [6,19]. This redundancy provides a complementary opportunity to test the consistency of interpretation.

Additionally, auxiliary tests of broadening and shift of the SA reference spectrum have been performed, in an attempt to compare, in the same vapour conditions, the volume spectra, and vW-sensitive SR spectra. For this purpose, auxiliary SA experiments (fig.3) were performed on a 100μm-thick Cs glass cell. This short thickness allows a high Cs density without a too strong absorption. In this regime, the SA experiments undergo an important broadening, that goes along a pressure shift. Interpretation of these pressure effects is discussed in the next section.

## 3. Data Analysis

As described in several of our previous works [1], our evaluation of the vW interaction relies on a comparison between the experimental (FM) SR lineshapes and a family of theoretical spectra [20, 21] depending on a single dimensionless parameter A that characterizes the vW interaction  [$A = 2 C_3 k^3 / \gamma$, with $k = 2\pi/\lambda$, and $C_3$ the vW coefficient governing the energy shift  $\Delta E = - C_3 z^{-3}$, with z the atom-surface distance). These calculated spectra undergo a combined shift and distortion relatively to the ideal vW-free (*i.e.* A=0) SR spectrum that would be obtained in similar conditions of atomic environment. Hence, it is a preliminary requirement to estimate the shift and broadening induced by the atom collision processes, such as those visibly appearing in the SA recordings at the higher temperatures (see fig. 3). Alternately, the pressure broadening and shift can be extracted from the SR spectrum by increasing the number of adjustable parameters in the fitting methods [6,7,14]. However, this usually increases uncertainties, and is sometimes paid by a limited radius of convergence of the adjustments. Moreover, in such an evaluation based on the SR experiment, the consistency of the pressure determination (*e.g.* linearity) has to be established.

### 3.1 Saturated Absorption and collision processes

Because of the limited linear absorption in a 100μm-thick cell, the SA spectra can be recorded in temperature conditions approaching those used in SR recordings. A limited number of temperature (pressure) is needed to get an estimate of the collision effects affecting the resonance. However, the ample overlaps between the hyperfine components, and even the vanishing resolution  of the hyperfine components for the higher Cs density, have imposed to use fitting methods to exploit the data. The lineshape of the individual hyperfine components can reasonably be assumed to be Lorentzian. The main difficulty is that the individual amplitude of the various components can *a priori* vary independently because SA is a nonlinear process. Moreover, the pressure shift and broadenings themselves could depend on the considered hyperfine component. This is why several fittings methods were compared [22], and notably: (i) a fully adjustable one, that includes a total of 16 parameters, namely a global "offset" parameter along with 3 parameters (position, amplitude and width) for each of the "5" individual components (there are 3 main hyperfine components and 3 additional Doppler crossover resonances, but the 3-5 crossover and the F'=4 principal component are intrinsically unresolved) ; (ii) one with individually varying amplitudes (because the saturation intensity strongly depends on the broadening itself), but for which the pressure effect imposes an identical pressure broadening and shift for all components; and (iii) a "global" one, for which in addition, the relative amplitudes are fixed (and determined by their relative values measured in low density conditions). As shown in figure 4, all these numerical methods yield comparable results, justifying not to discriminate between the hyperfine components. Indeed, it is essentially when the h.f.s is well-resolved that individual behaviours can be expected. In our present experimental conditions, the h.f.s. is already washed out by the pressure broadening. The accuracy of the results, especially for frequency shifts that amounts to less than 10 MHz, is anyhow affected by the laser jitter and lack of reproducibility in small-scale frequency scan. Note that convergence of the numerical fitting is more easily obtained for moderate temperature (*i.e.* hyperfine structure partially resolved), than for the higher temperature when the pressure broadening smears out the hyperfine structure. From this SA analysis, we can extract an evaluation for the pressure broadening (~ 180 MHz/Torr) and shift (~ - 60 MHz/Torr). However, the high-level of uncertainty on these data, notably related to the pressure uncertainty (as extrapolated from a temperature measured with a few degrees uncertainty) (see *e.g.* [14]), would make it highly hazardous to convert these estimates into a known correction imposed for the analysis of SR spectra. Rather, because the pressure shift remains in most cases comparable to the laser jitter, the estimated SA pressure shift can be included in the consistency check when fitting the SR spectra. Similarly, the pressure uncertainty when comparing SR and SA experiments at a same nominal temperature explains why it is not convenient in the SR fitting to impose the pressure broadening as estimated from the SA experiment.

### 3.2 Analysis of Selective Reflection spectra

As justified by our results concerning the SA spectra, our fittings for the SR spectra generally assume an adjustable width $\gamma$ and shift $\Delta$ identical for all hyperfine components. An identical vW interaction, governed by a single $C_3$ coefficient [23] is also assumed. Moreover, and owing to the linear nature of SR spectroscopy, the SR spectra can be fitted by taking as the relative amplitudes of the hyperfine components those imposed by the angular factors for transition probabilities [7]. The SR fittings hence result from the optimal least-square adjustments obtained, for a series of A values, through simple homothetic changes relatively to the theoretical shape. Figure 5 illustrates, for a given SR spectrum, a choice of A values producing acceptable (or nearly acceptable) fits. The experimental spectrum being here recorded at a temperature for which the hyperfine structure is hardly resolved, and for which a possible pressure shift has to be considered, the *a priori* position of the individual SR hyperfine components is determined with respect to the corresponding positions in the SA spectrum [24], and after addition of a pressure shift identical for all h.f. components. For a more systematic determination of the range of acceptable A parameters, we plot as in fig. 6a the total error value for the optimal fit for a given A value, as a function of the A value. Such a plot is especially of interest when the pressure broadening washes out details in the SR spectrum, so that numerous fits, for a large range of A values, can produce resembling spectra. As shown in fig. 6b, it appears that in the considered strong vW regime, the fitted $\gamma$ value itself is strongly dependent upon the choice of the A parameter - see (fig 6b)- . This behaviour is at the opposite of the one in weak (A<<1) vW regime, when the determination of the width $\gamma$ is nearly A-independent [14]. As can be seen in fig. 6c, such a behaviour tends to compensate for the uncertainty on the optimal A parameter, hence reducing the overall uncertainty on A$\gamma$ or $C_3$.

Through the development of an automated numerical method of fitting [22], we could perform a more systematic determination of the optimal vW value fitting the experiments, increasing the confidence in the $C_3$ determination. Instead of optimizing a freely adjustable "pressure shift" -ultimately expected to be proportional to pressure- , we have systematically compared the fitting values assuming a given shift between the SA reference and the SR cell. In this exploration, the given shift is typically changed by a few MHz increments. As shown in fig 7a, the region of optimal A value drifts under this hypothetic shift, while correlatively the cloud of acceptable values of $\gamma$ remains horizontal. The estimated width $\gamma$ hence appears independent of the hypothetic pressure shift. Fig. 7c shows that this method provides a narrow range of acceptable values of A$\gamma$ (*i.e.* of $C_3$). In addition, the amplitude of the SR spectrum, after normalization relatively to the theoretical vW curve, provides in the principle a measurement of the number of interacting atoms [6,25]. Note that this "signal amplitude" remains nearly unaffected by the precise choice of the hypothetic shift, provided that one chooses accordingly the optimal A value.

The major results of our analysis are shown in fig. 8, where the results of the SR fittings, including the uncertainties resulting from the range of acceptable "pressure" shifts, are synthesized as a function of Cs pressure. Remarkably, the $C_3$ value extracted from the various fittings is independent of the pressure. The value to be retained, $C_3 = 73 \pm 10$ kHz.$\mu m^3$ , is noticeably larger than the one found for the second resonance line of Cs ($\sim$ 10-20 kHz.$\mu m^3$)[6,7], and comparable in magnitude -but here for an attraction- to the "giant" repulsion ($\sim$ -160 kHz.$\mu m^3$) [8] obtained in a situation of resonant coupling. Note that part of the uncertainty in this $C_3$ value can be traced back to the uncertainty in the frequency calibration for small frequency scans. Moreover, the extrapolated widths, shifts, and amplitude, evolve linearly with the Cs pressure. This is a major *a posteriori* justification of the pressure origin of the *ad hoc* shift introduced in the fitting method. One approximately extracts a 280 MHz/Torr dependence for $\gamma$, and a - 90   MHz/Torr value for $\Delta$. Also, these estimated pressure effects are compatible with the estimates extracted from the SA experiment. Note that the large uncertainty in the absolute pressure calibration (a 5°C error in the temperature measurement may modify the pressure scale by 40 %), makes the SR and SA comparison imprecise, but does not affect the internal coherence of measurements mainly sensitive to temperature variations. also independent of the considered hyperfine initial level $6S_{1/2}$(F=3), or $6S_{1/2}$(F=4). Also, the extracted $C_3$ values are found to be the same (within 5 %) for the F=4$\rightarrow$F'={3,4,5} and F=3$\rightarrow$F'={2,3,4} components [26].

## 4. Discussion and comparison with the theoretical predictions

The $C_3$ measured coefficient in SR spectroscopy is related to the dissimilar vW attraction exerted on the final ($|$ f$>$) and initial ($|$ i$>$)  states (*i.e.* $C_3 = C_3(|$ f$>$) - $C_3(|$ i$>$)]. Because the VW attraction quickly increases with the atomic excitation, and is roughly connected with the atomic level polarizability [3], one has in our case $C_3(6S_{1/2}) << C_3(8P_{3/2})$ . With $C_3(6S_{1/2}) \leq 2$kHz.$\mu m^3$, one simply deduces that the measured $C_3 = 73 \pm 10$ kHz.$\mu m^3$ totally applies to $C_3(8P_{3/2})$. In an evaluation entirely analogous to previous ones, Table 1 shows how the various dipole couplings relevant for the $8P_{3/2}$ level of Cs respectively contribute to the vW attraction coefficient $C_3(8P_{3/2})$. The different columns are relevant for an interface either with an ideal reflector, either with a sapphire interface. Because there are no coincidences between sapphire surface resonances, and the dipole transitions reaching  the $8P_{3/2}$ level of Cs, the dielectric coefficients are not dramatically sensitive to uncertainties affecting the sapphire properties or orientation [8,27].

It is remarkable that in spite of the expected accuracy of the theoretical prediction, the measured $C_3$ value appears sensitively higher - by a factor 1.5- than the expected one ( $\sim 45$ kHz.$\mu m^3$), and hence closer to the value predicted for an ideal reflector than with the dielectric image factor taken into account. We do not have at the present time a fully satisfactory explanation of this fact. The laser jitter, although leading to a lineshape broadening, susceptible to modify the family of theoretical vW lineshapes themselves does not appear susceptible to modify sensitively our $C_3$ value estimate. However, Table 1 shows that more than 60% of the $C_3$ value originates in the virtual absorption, located in the far IR range, towards the 7 D level. Even for a material like sapphire that does not exhibit a surface resonance around this 36-39$\mu m$ range, the theoretical modelling of a vacuum at zero temperature should no longer hold. Rather, modifications of the vW interaction are to be expected with our hot sapphire surface (for the overheated window, we have currently T $\sim$450-500K). The evaluation of this non resonant thermal effect appears tractable after recent theoretical developments [9]. Also, straightforward extrapolations based upon the sapphire properties as measured at room temperature could be invalid, because the sapphire dielectric susceptibility itself could be modified in the thermal IR range, when high temperatures are considered. It is interesting to note that if a thermal influence on the vW surface interaction has never been reported, the high precision experiments on Cs Rydberg levels [3] had been performed at room temperature, with no noticeable deviation in the interaction strength, relatively to the T=0 prediction, although the dominant dipole couplings lie in a very remote IR range, Meanwhile, these experiments on Rydberg atoms were not performed in a genuine half-space but between two walls, so that, as demonstrated for the real transfer induced by the surface thermal excitation [28], the sub-wavelength cavity may prevent the influence of the thermal excitation on the vW interaction.

Finally, the remaining disagreement between our experimental findings, and the theoretical prediction (as evaluated in a T=0 approach, and assuming for sapphire similar properties than at room temperature), appears because of our refined accuracy in the interpretation of SR data. Until now [1, 6-8, 13,14], the agreement between the theoretical predictions and the experiments had remained within error bars that were much larger. If part of our present experimental uncertainty can even be attributed to defects in the frequency scale calibration, a 10 % accuracy could be envisioned with a better control of the laser frequency scan. From the extreme quality of the overlap between the experimental spectra, and the complex fitting curves, requiring a restricted number of parameters, one may feel that one holds a clear evidence of a $z^{-3}$ behaviour. In spite of that, one cannot totally exclude the influence of phenomena related with the surface state itself that are not taken into account. In particular the effect of the local roughness of the surface (i.e. z would be replaced by a locally-averaged <z> value) could become more important than in previous situations, owing to the relatively short wavelength that we use, implying a more accurate spatial resolution [5]. Also, residual electric fields at the surface, as possibly existing on a dielectric surface, or as induced by residual Cs layers, are susceptible to induce a Stark shift, to which Cs (8P) could be more sensitive than in most previous experiments, owing to a higher degree of electronic excitation.

### *Acknowledgements*

P.C.S. Segundo acknowledges the support of Brazilian CAPES (BEX 0741/01-9) for his stay in France. We also acknowledge the participation and help for some stages of the experiments of Paulo Valente -in the frame of an ECOS French-Uruguayan programme- , of Gabriel Dutier, and of Alexander Yarovitski. We also acknowledge various discussions on these topics with S. Saltiel (French Bulgarian RILA cooperation) and with J.R.R. Leite (French-Brazilian CAPES-COFECUB 456/04). This work is part of the FASTnet project, partially supported by the European Union (contract HPRN 2002-00304).

**Notes and References**

**Figure captions**

Figure 1 : Scheme of the experimental set-up, with the atomic energy levels relevant for the $6S_{1/2} \rightarrow 8P_{3/2}$ transition of Cs. The output beam of the isolator needs to be optically isolated - "O.I." -, and cleaned-up with a spatial filter.

Figure 2 : Experimental (FM) SR lineshapes on the F=4 $\rightarrow$F'={3,4,5} component of $6S_{1/2} \rightarrow 8P_{3/2}$ transition of Cs as recorded at : (a) T= 172°C, or ~ $6.10^{14}$ at/cm$^3$ ; (b) T= 181°C, or ~ $9.10^{14}$ at/cm$^3$ ; (c) T=227°C, or ~ $5.10^{15}$ at/cm$^3$. The SA reference (on the bottom) is an overlap of the SA spectra recorded simultaneously with SR spectra in conditions (a) to (c). The continuous lines (colour on line) superimposed to the SR spectra are fitting lineshapes. The signal amplitude in the FM technique depends on the amplitude of the FM modulation; it is here chosen to optimize - but not to broaden- the narrower signal (a) : this explains that the amplitude of the FM signal is not notably larger in (b) and (c), in spite of an increased number of atoms.

Figure 3 : SA spectra recorded in the 100μm-thick cell at various high temperatures: (a) T= 155°C, or ~ $3.10^{14}$ at/cm$^3$ ; (b) T= 180°C, or ~ $8.10^{14}$ at/cm$^3$ ; (c) T= 214°C, or ~ $3.10^{15}$ at/cm$^3$ ; (d) SA spectrum in the macroscopic reference cell at a low-temperature (T=80°C or ~ $4.10^{12}$ at/cm$^3$)

Figure 4 : Pressure broadening (γ) and pressure shift (Δ) as extracted from the SA measurements. The results of the various fitting methods (square: fixed relative h.f. amplitudes; circle: independent h.f. amplitudes but a single pressure effect; triangle: independent amplitudes and pressure effects for each h.f. component), plotted here for the pressure shift, exhibit only negligible differences (the distribution of parameters for the "independent" h. f. component is quite narrow ~ 10% when convergence can be obtained). The uncertainty in the pressure scale is on the order of 40 % , corresponding to an absolute 5°C temperature uncertainty.

Figure 5 : Comparison of optimally fitting lineshapes for the experimental SA spectrum of fig. 2b. For A =10, 14, 18, the optimal fittings yield respectively Aγ = 556, 587, 616 MHz , *i.e.* respectively C$_3$= 65, 69, 73 kHz.μm$^3$, and γ = 56, 42, 34 MHz. The fittings here allow for a frequency shift (to be attributed to a pressure shift) relatively to the SA reference, optimized to a - 14 MHz value for the 3 proposed fittings

Figure 6 : Optimal adjusted fitting parameters as a function of the choice of the A value, for the experimental lineshape of fig 2b (see also fig. 5): (a) plot of the error value (mean square

error); (b) Plot of the γ fitting value; (c) Plot of the Aγ fitting value. Each fitting is optimized independently of the SA reference (*i.e.* the extracted pressure shift, although modest, varies with the choice of the A value)

Figure 7 :  Same as figure 6, but with an imposed pressure shift $\Delta$ (as indicated).

Figure 8 :  Synthesis of the optimal fittings for various experimental conditions, as a function of the pressure for the F=4 h.f.s. component  of Cs($6S_{1/2}$) : (a) Resulting value of Aγ. The average   Aγ value is equivalent to $C_3$=73 kHz.$\mu m^3$, (b) Plot of the estimated pressure broadening, (c) Plot of the estimated pressure shift, (d) Plot of the estimated amplitude.

**Table caption**

Table 1 : Table indicating the contribution to the $C_3$ coefficient for Cs($8P_{3/2}$) of each virtual transition to the $C_3$ coefficient for Cs($8P_{3/2}$), on the model of ref. 6, with the relevant transition wavelength (sign minus for a virtual emission), the transition probability $A_{ij}$, the relevant geometrical coefficient. The column $C_3^M$ is for an ideal reflector, the column $C_3^S$ is for a sapphire window, with the intermediate column $r_S$ the relevant coefficient of dielectric image for sapphire.

Figure 1

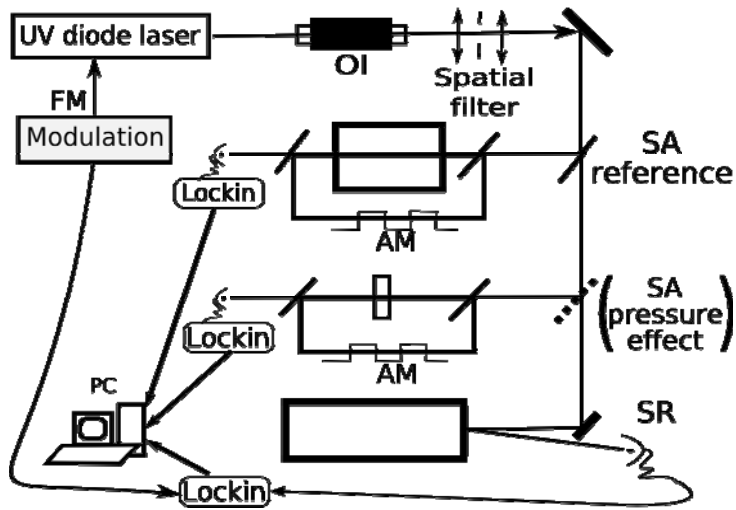

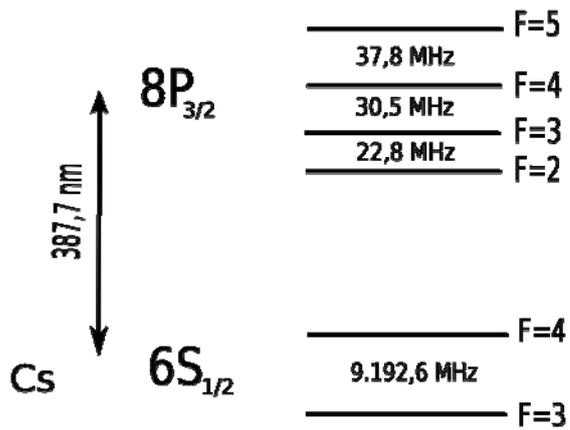

Figure 2

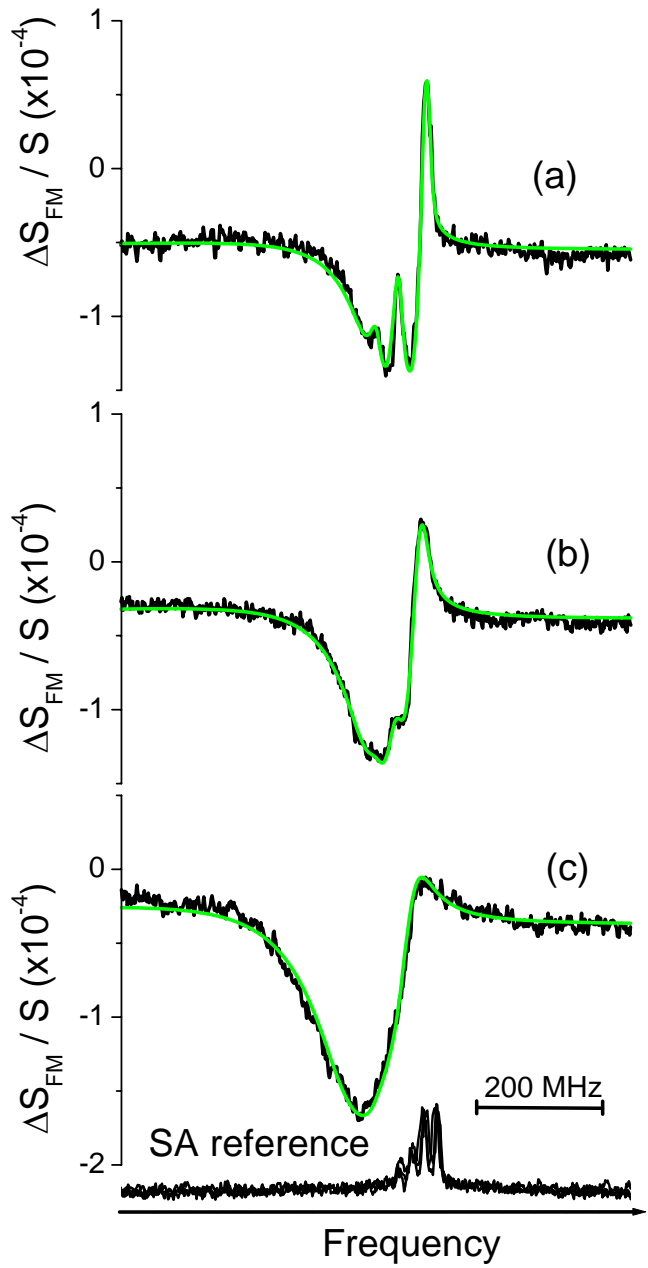

Figure 3

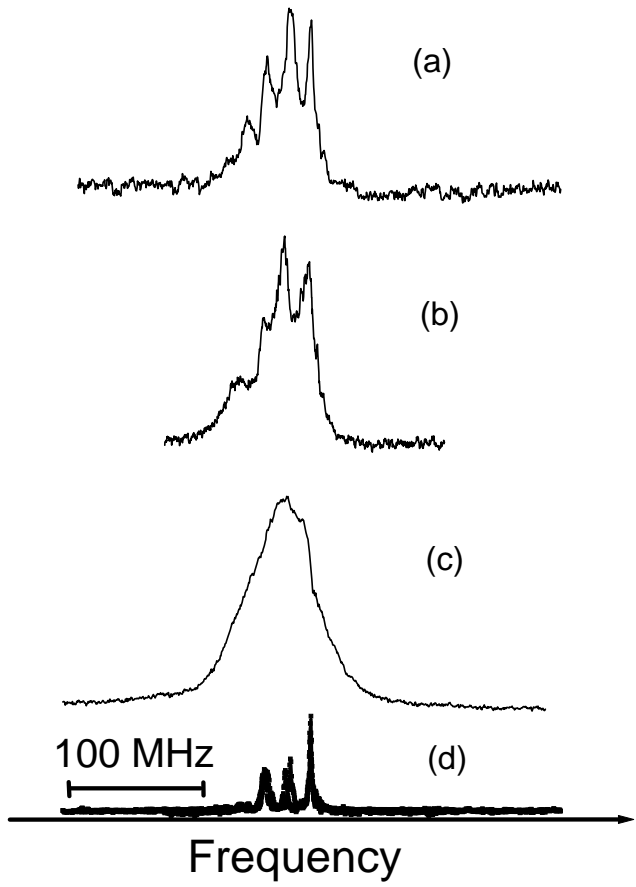

(a)

(b)

(c)

100 MHz

(d)

Frequency

Figure 4

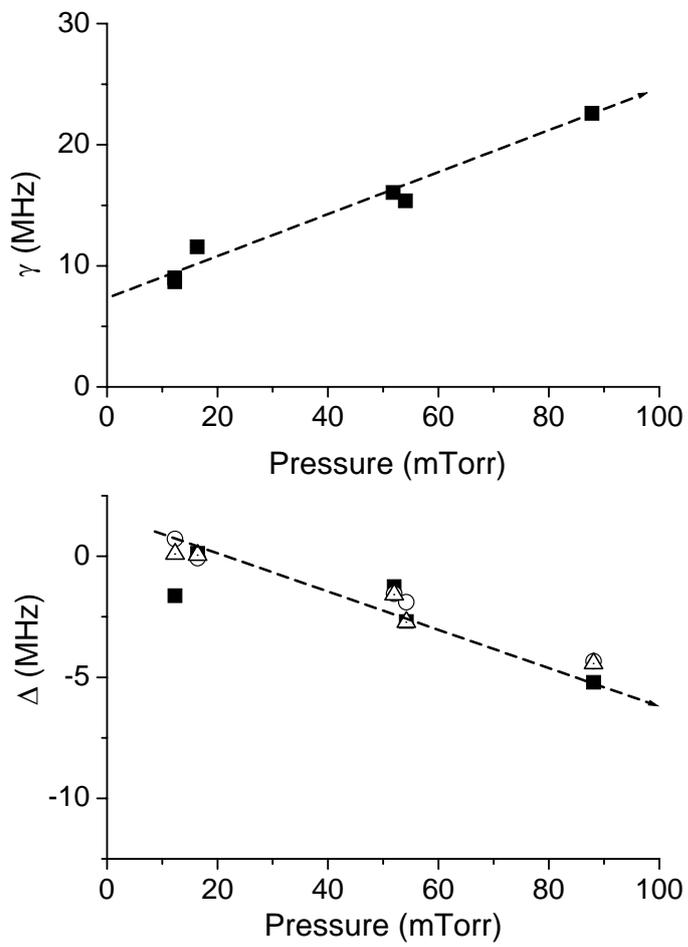

Figure 5

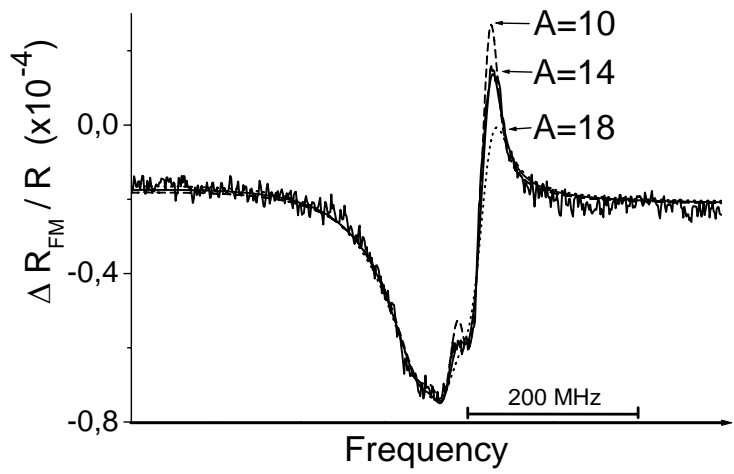

Figure 6

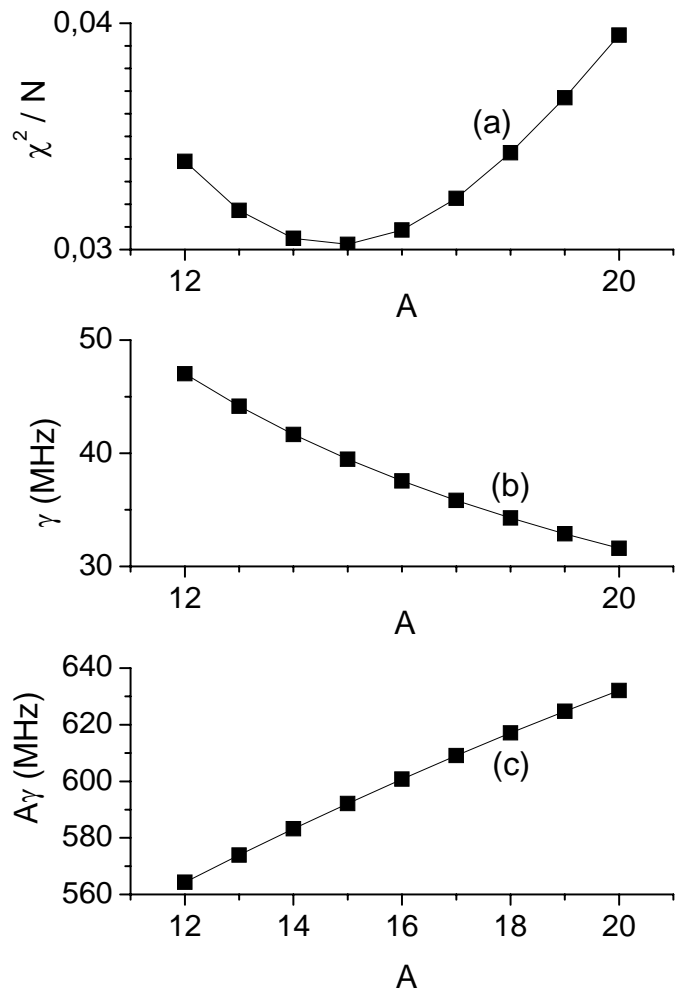

Figure 7

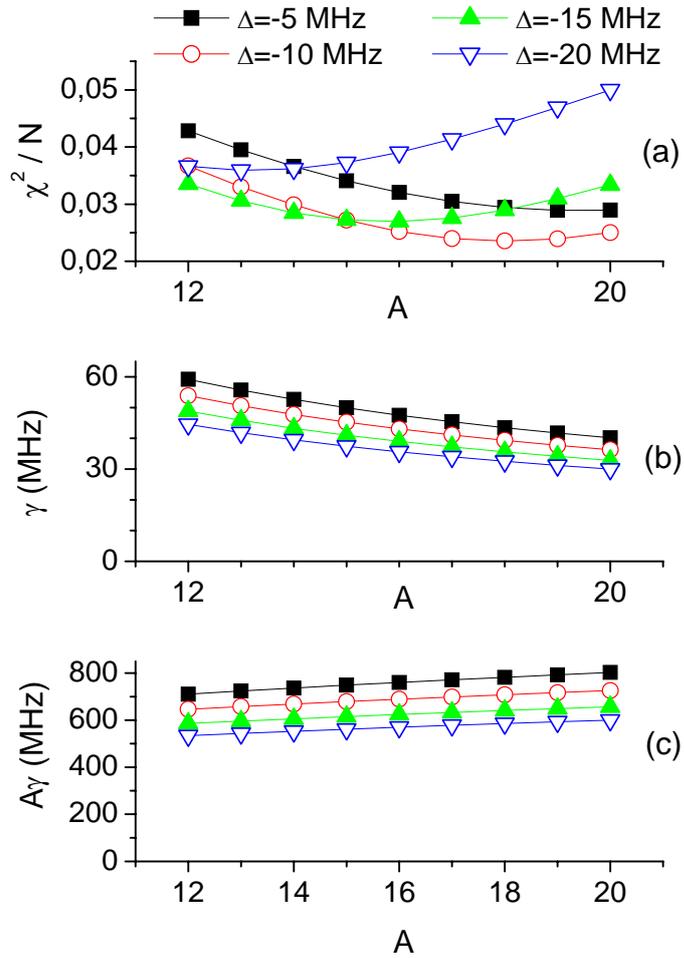

Figure 8

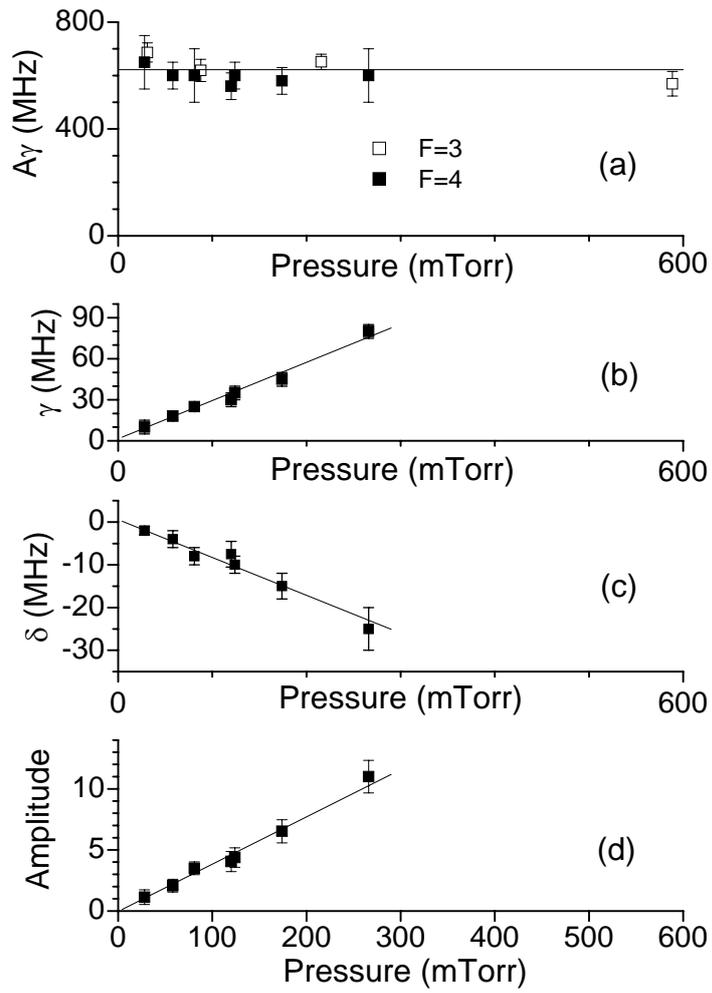

Table 1

.

| Level | λ (μm) | $A_{ij}$ ($10^4$ s$^{-1}$) | $\dfrac{2J_> + 1}{2J_0 + 1}$ | $C_3^M$(kHz.μm$^3$) | $r_S$ | $C_3^S$(kHz.μm$^3$) |
|---|---|---|---|---|---|---|
| 5D$_{3/2}$ | -0.89 | 37.9 | 1 | <0.01 | 0.506 | <0.001 |
| 5D$_{5/2}$ | -0.89 | 37.9 | 1 | 0.01 | 0.506 | 0.005 |
| 6D$_{3/2}$ | -3.12 | 61.8 | 1 | 0.08 | 0.421 | 0.032 |
| 6D$_{5/2}$ | -3.16 | 61.8 | 1 | 0.71 | 0.419 | 0.296 |
| **7D$_{3/2}$** | **39.05** | **1.34** | **1** | **5.32** | **0.749** | **3.980** |
| **7D$_{5/2}$** | **36.09** | **1.34** | **3/2** | **37.78** | **0.744** | **28.111** |
| 8D$_{3/2}$ | 4.95 | 51.6 | 1 | 0.42 | 0.602 | 0.252 |
| **8D$_{5/2}$** | **4.92** | **51.6** | **3/2** | **3.70** | **0.602** | **2.226** |
| 9D$_{3/2}$ | 3.29 | 37.4 | 1 | 0.09 | 0.578 | 0.051 |
| 9D$_{5/2}$ | 3.28 | 37.4 | 3/2 | 0.80 | 0.578 | 0.460 |
| 10D$_{3/2}$ | 2.72 | 25.9 | 1 | 0.03 | 0.568 | 0.020 |
| 10D$_{5/2}$ | 2.72 | 25.9 | 3/2 | 0.31 | 0.568 | 0.117 |
| 11D$_{3/2}$ | 2.44 | 18.2 | 1 | 0.02 | 0.530 | 0.010 |
| 11D$_{5/2}$ | 2.43 | 18.2 | 3/2 | 0.16 | 0.553 | 0.089 |
| 12D$_{3/2}$ | 2.27 | 13.3 | 1 | 0.01 | 0.560 | 0.006 |
| 12D$_{5/2}$ | 2.27 | 13.3 | 3/2 | 0.09 | 0.560 | 0.052 |
| | | | | | | |
| 6S$_{1/2}$ | -0.39 | 125 | 1 | 0.03 | 0.534 | 0.002 |
| 7S$_{1/2}$ | -1.38 | 56.6 | 1 | 0.31 | 0.488 | 0.029 |
| **8S$_{1/2}$** | **-6.78** | **96.5** | **1** | **12.07** | **0.158** | **1.906** |
| **9S$_{1/2}$** | **8.94** | **122** | **½** | **11.63** | **0.643** | **7.477** |
| 10S$_{1/2}$ | 3.99 | 45.1 | ½ | 0.38 | 0.588 | 0.225 |
| 11S$_{1/2}$ | 3.00 | 25.3 | ½ | 0.09 | 0.573 | 0.052 |
| 12S$_{1/2}$ | 2.58 | 15.4 | ½ | 0.04 | 0.566 | 0.020 |
| Total | | | | 73.79 | | 45.47 |